# Delta-doped β-Ga$_2$O$_3$ Films With Low FWHM Charge Profile Grown By Metalorganic Vapor-Phase Epitaxy

Praneeth Ranga,[1] Arkka Bhattacharyya[1], Adrian Chmielewski[2], Saurav Roy[1], Nasim Alem[2] and Sriram Krishnamoorthy[1]

[1]*Department of Electrical and Computer Engineering, University of Utah, Salt Lake City, 84112, USA*

[2] *Department of Materials Science and Engineering, Pennsylvania State University, University Park, State College, PA 16802, USA*

## Abstract

We report on low-temperature MOVPE growth of silicon delta-doped β-Ga$_2$O$_3$ films with low FWHM. The as-grown films are characterized using Secondary-ion mass spectroscopy, Capacitance-Voltage and Hall techniques. SIMS measurements show that surface segregation is the chief cause of large FWHM in MOVPE-grown films. The surface segregation coefficient (R) is observed to reduce with reduction in the growth temperature. Films grown at 600 °C show an electron concentration of 9.7 x 10$^{12}$ cm$^{-2}$ and a FWHM of 3.2 nm. High resolution scanning/transmission electron microscopy of the epitaxial film did not reveal any significant observable degradation in crystal quality of the delta sheet and surrounding regions. Hall measurements of delta-doped film on Fe-doped substrate showed a sheet charge density of 6.1 x 10$^{12}$ cm$^{-2}$ and carrier mobility of 83 cm$^2$/V. s. Realization of sharp delta doping profiles in MOVPE-grown β-Ga$_2$O$_3$ is promising for high performance device applications.

Ultrawide bandgap β-Ga$_2$O$_3$ is a promising material for high-power and high-frequency applications. The high critical breakdown field enables fabrication of devices with thinner active region leading to a significant reduction in on-resistance compared to other semiconductors[1]. Additionally, the availability of high-quality, large area, single-crystal substrates and controllable n-type conductivity makes β-Ga$_2$O$_3$ a potential competitor to existing power semiconductors[2]. Growth of β-Ga$_2$O$_3$ and n-type doping has already been studied utilizing a variety of growth techniques[3–11]. Significant progress has been made in growth and fabrication of β-Ga$_2$O$_3$ based power devices. Critical breakdown field strength up to 5.7 MV/cm has already been realized in BTO/β-Ga$_2$O$_3$ heterostructure Schottky diodes[12]. Breakdown voltage of 8 kV has been demonstrated in lateral β-Ga$_2$O$_3$ MOSFET, indicating great promise for β-Ga$_2$O$_3$ based power electronics[13].

Most of the β-Ga$_2$O$_3$ devices are based on uniformly-doped channels, which limits the performance and design space of power and RF devices. Channels with abrupt doping profiles can offer improved performance in terms of on-current, breakdown voltage and higher transconductance than a uniformly doped channel[14]. Delta-doped channels are expected to have reduced scattering with the ionized donor atoms due to the wavefunction spreading into the neighboring undoped regions, which can lead to an increase in carrier mobility[15]. High-density MBE(Molecular beam epitaxy)-grown delta-doped layers with sheet charge up to $1 \times 10^{13}$ cm$^{-2}$ and room temperature mobility of 75 cm$^2$/V. s has been realized recently[16]. By coupling multiple delta sheets of high sheet charge, a high carrier mobility of 83 cm$^2$/V.s has been realized at a sheet charge density of $2.4 \times 10^{14}$ cm$^{-2}$ using MBE-grown β-Ga$_2$O$_3$ films[17]. By shrinking the gate length aggressively (~250 nm) and realizing low resistance ohmic contacts, delta-doped FET with high current density and cut-off frequency (27 GHz) have been realized using MBE[16,18]. However, most of the research till date is focused on MBE growth of delta-doped and modulation-doped films. Recent reports of record high mobility in MOVPE-grown β-Ga$_2$O$_3$ films indicate the promise of MOVPE (Metalorganic vapor-phase epitaxy) technique for fabrication of high power β-Ga$_2$O$_3$ devices[3,8,19,20]. High-quality epilayers with low FWHM (full width at half maximum) and high-density sheet charge are required to realize state-of-the-art device performance using MOVPE material.

In addition to use as a channel layer, study of delta doping is critical for attaining high density 2DEG(two-dimensional electron gas) at β-(Al$_x$Ga$_{1-x}$)$_2$O$_3$/ β-Ga$_2$O$_3$ heterostructures[21]. Theoretical predictions show that 2DEG mobility values exceeding the limit of bulk β-Ga$_2$O$_3$ can be attained by realizing a high-density 2DEG at the heterointerface[22]. To realize a high density 2DEG, it is critical to obtain a sharp high-density sheet charge in the β-(Al$_x$Ga$_{1-x}$)$_2$O$_3$ barrier layer. Additionally, barriers with higher Al content are important for confining a high-density 2DEG. MOVPE growth technique is promising for growth of high-composition β-(Al$_x$Ga$_{1-x}$)$_2$O$_3$ with Al composition up to 40 % has been realized in MOVPE[23]. N-type doping of β-(Al$_x$Ga$_{1-x}$)$_2$O$_3$ is predicted to be feasible for Al composition up to 0.8[24], experimentally n-type doping of β-(Al$_x$Ga$_{1-x}$)$_2$O$_3$ films has already been reported by multiple groups[23,25]. Insofar, delta doping in β-Ga$_2$O$_3$ has only been studied by MBE and MOCVD techniques[16,26]. However, sharp delta sheets with low FWHM has only been realized using MBE technique[16]. In this work, we have studied growth of MOVPE-grown delta-doped films and realized sharp doping profiles with narrow FWHM. Using SIMS (secondary ion mass spectroscopy) analysis, silicon surface segregation is identified as the key contributor to large FWHM of the delta sheets. By lowering the growth temperature from 810 °C to 600 °C, delta sheet charge of $9.7 \times 10^{12}$ cm$^{-2}$ and FWHM of 3.2 nm is realized. Hall measurements of β-Ga$_2$O$_3$ films on Fe-doped substrates indicate that the carrier mobility can be improved by utilizing coupled multiple delta-doped layers.

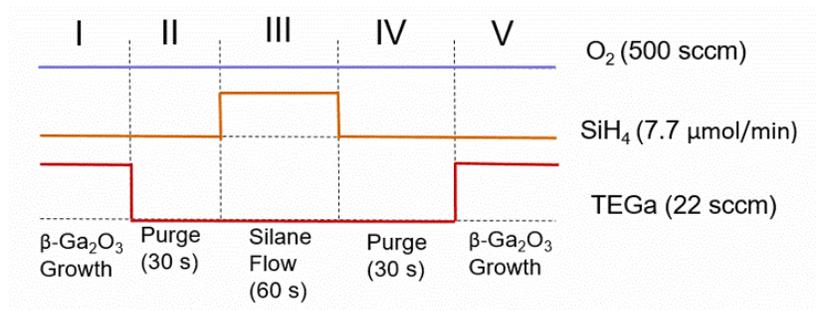

**Fig 1.** Timing diagram utilized for realizing MOVPE-grown delta-doped β-Ga$_2$O$_3$ films

Growth of β-Ga$_2$O$_3$ is performed using Agnitron Agilis MOVPE reactor with Triethyl Gallium (TEGa), Trimethyl aluminum (TMAl) and oxygen (O$_2$) as precursors and Argon as carrier gas. Diluted Silane (SiH$_4$) is used as silicon source for achieving controllable n-type doping. Delta doping of β-Ga$_2$O$_3$ is achieved through a growth interruption-based process where the flow of TEGa is stopped and silane is flowed into the chamber. A pre- and post-purge step (30s each) is performed to remove any unreacted precursors from the growth chamber. For achieving delta doping, silane is supplied to the chamber under an oxygen ambient (Fig. 1). UID β-Ga$_2$O$_3$ is grown before and after the delta doped layer, which acts as a buffer and cap layer respectively. The density of silicon delta sheet is controlled by tuning silane flow, silane time, pre- and post-purge steps. Such growth interruption-based approach has already been utilized for delta doping of MOVPE-grown GaN films[27,28]. To obtain reliable CV (Capacitance-voltage) measurements, the growths are performed on a Sn-doped (010) β-Ga$_2$O$_3$ substrates for the ease of Ohmic contact formation. Post growth Ti/Au (50/50 nm) thick contacts are deposited via sputtering to obtain backside ohmic contact to the substrate. Ni/Au (50/50 nm) contacts are deposited via ebeam evaporation to obtain Schottky contact to the top layer after patterning using photolithography. CV measurements are performed on the fabricated diodes to obtain charge density and distribution of the silicon delta sheet charge. Epitaxial films with similar growth conditions are also grown on Fe-doped β-Ga$_2$O$_3$ substrates for hall characterization. Additionally, multiple delta sheets with varying growth conditions are also grown on Fe-doped substrates to measure the SIMS silicon density as a function of growth conditions.

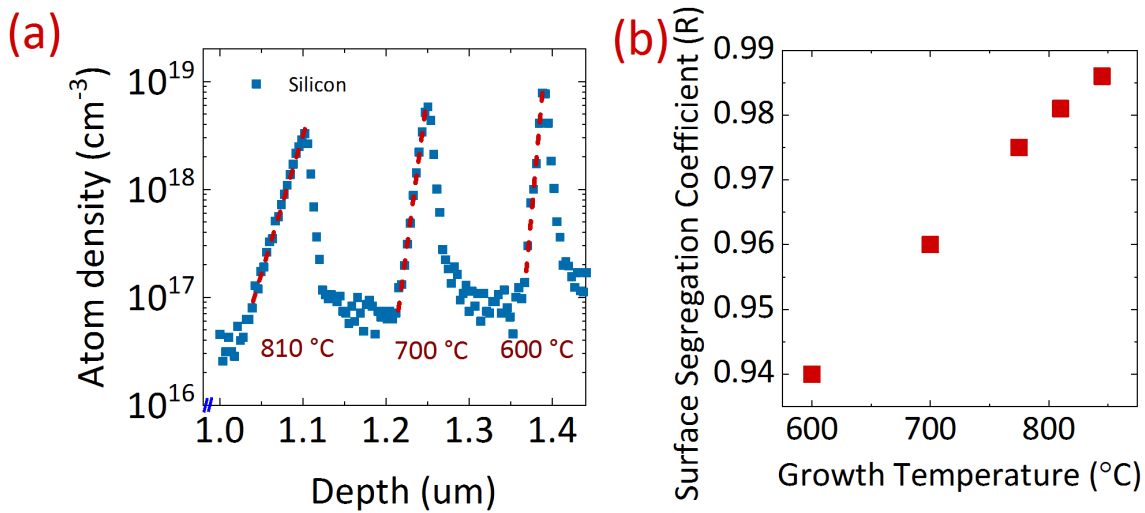

**Fig.2** (a) SIMS profiles of Si delta-doped layers grown at 600 °C, 700 °C and 810 °C. (b) Variation of the surface segregation coefficient (R) with growth temperature.

In addition to electrical characterization, SIMS measurements are also performed to understand the actual silicon atom density and spread of the silicon atoms. Silicon delta sheets are grown at different growth temperatures while keeping other process parameters the same (silane flow time – 60s, pre-purge – 30s, silane flow – 7.7 nmol/min, post-purge- 30s). Silicon delta doped layers are grown with UID $Ga_2O_3$ spacer layers (~ 150 nm). Carbon and hydrogen concentrations were below/close to detection limit, indicating low concentration of external impurities (see supplementary data) even at 600 °C growth temperature. The SIMS profiles for delta sheets grown at 810 °C, 700 °C and 600 °C are plotted in figure. 2(a). The integrated Si atom density has a weak dependence on growth temperature. The silicon density reduced from $1.1 \times 10^{13}$ $cm^{-2}$ to $8.7 \times 10^{12}$ $cm^{-2}$ with increase in growth temperature (600 °C – 810 °C), this effect could be attributed to increased Si desorption at higher temperatures. Delta sheets grown at higher growth temperatures show more spread in the silicon profiles compared to other films with similar sheet concentration. Further analysis revealed that the tail of the Si profile can be modeled using the surface segregation model. Surface segregation of dopant impurities has been observed in epitaxial growth of compound semiconductors[29–31]. Recently, Fe surface segregation has been observed in MBE growth of β-$Ga_2O_3$[31]. The value of surface segregation coefficient (R) indicates the amount of surface riding of impurity atoms, higher the value of R lower the silicon incorporation. By calculating an initial surface coverage ($\theta_0$), the surface coverage ($\theta_n$) and sheet concentration [$Si_n$] can be determined for each monolayer 'n' in the film. The value of R can be calculated using R = $([Si]_N /[Si]_1 )^{1/(N-1)}$, where $[Si]_1$ and $[Si]_N$ are the sheet concentrations for the first and nth layer respectively[31]. Using the formalism in literature[29,31], the surface segregation coefficient (R) is calculated for the Si profiles grown at different temperatures. The calculated silicon profile assuming a constant value of R is indicated by the dashed line in Fig.2(a). With reduction in growth temperature, SIMS FWHM of the delta sheet reduces from 21 nm to 11 nm from 810 °C to 600 °C. The calculated model and the experimental results agree very well; this shows that surface segregation is the dominant cause of spread in silicon profiles. However, the mechanism behind the backward silicon tail (towards the substrate) of silicon is still unclear. Si diffusion cannot explain the dopant profile since the diffusion of silicon is prominent only at temperatures much higher than the MOVPE growth temperature, as reported from ion implantation experiments[32,33] (~ 1100 °C). However, the silicon atom diffusion could be enhanced by vacancies and interstitials near the growth surface. Further experiments are needed to identify the cause of this phenomenon. To obtain delta sheets with low FWHM, reducing the segregation coefficient R is very critical. Hence, it is important to understand how R varies as a function of temperature. Fig. 2(b) shows the variation of R with growth temperature, R reduces proportionally with reduction in growth temperature. Similar temperature dependence of R has been observed in MBE growth of InGaAs, GaN and β-$Ga_2O_3$[29,31,34]. This suggests that for obtaining sharp dopant profiles, the growth temperature should be reduced to the lowest allowable value.

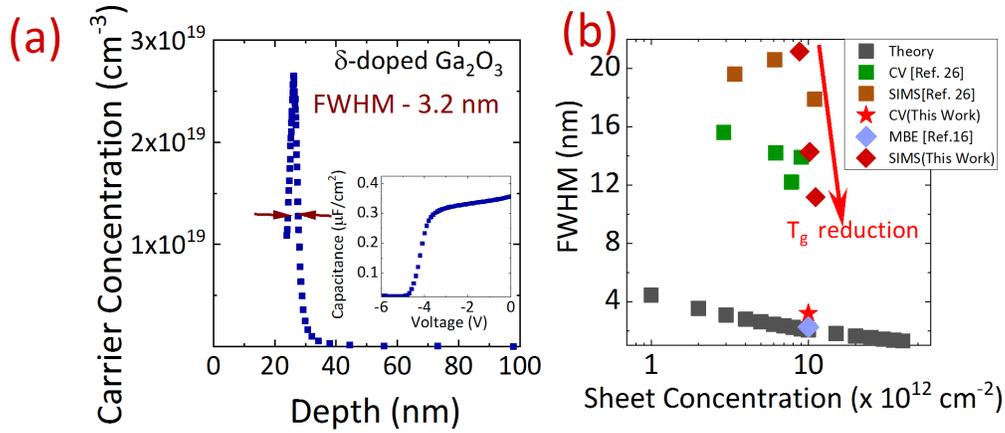

**Fig.3** (a) CV measured charge profile of delta-doped β-$Ga_2O_3$ epilayers grown at 600 °C with inset of CV data (b) Plot of extracted FWHM (SIMS and CV) of Si delta sheet as a function of sheet charge density

CV measurements were performed to understand the donor activation of the silicon delta sheet. A single delta sheet identical to the growth in Fig.2(600 °C) is grown on a Sn-doped substrate with 25 nm spacer and 450 nm buffer layer. CV measurements showed a sheet charge density of 9.7 x $10^{12}$ $cm^{-2}$ and FWHM of 3.2 nm (Fig. 3(a)). In a uniformly doped semiconductor, the resolution of the CV profile is limited by the debye length. In a degenerate semiconductor with quantum confinement, the CV resolution is limited by the spread of the electron wavefunction. This places a lower limit on the FWHM measurement of the silicon delta sheet[35]. For an ideal delta sheet with zero spread in the donor profile, the FWHM equals $2(7/5)^{0.5} (4\varepsilon\hbar^2/9e^2N^{2D}m^*)^{(1/3)}$. Where ℏ is the reduced Planck's constant, ε is the dielectric constant of (010) oriented β-$Ga_2O_3$ (10), m* is the conduction band effective mass in β-$Ga_2O_3$ (0.28 $m_0$), e is charge of the electron and $N^{2D}$ is the electron sheet charge density. The measured FWHM of the silicon delta sheet is plotted in fig.3(b), FWHM measured by SIMS and CV is plotted as a function of sheet charge density. Films grown at higher temperatures have a relatively large FWHM values[26]. The SIMS measured FWHM reduces from 20 nm to 10 nm with reduction in growth temperature from 810 °C to 600 °C. This is attributed to reduction in surface segregation with growth temperature as explained before. Similarly, CV measured FWHM of low temperature films(3.2 nm) is significantly smaller than the FWHM of samples grown at 810 °C (12 – 15 nm)[26]. Additionally, the FWHM of MOVPE silicon delta sheet is comparable to MBE grown films[16], which is promising for MOVPE grown β-$Ga_2O_3$ films. Lower FWHM measured by CV compared to SIMS is attributed to the quantum confinement effect.

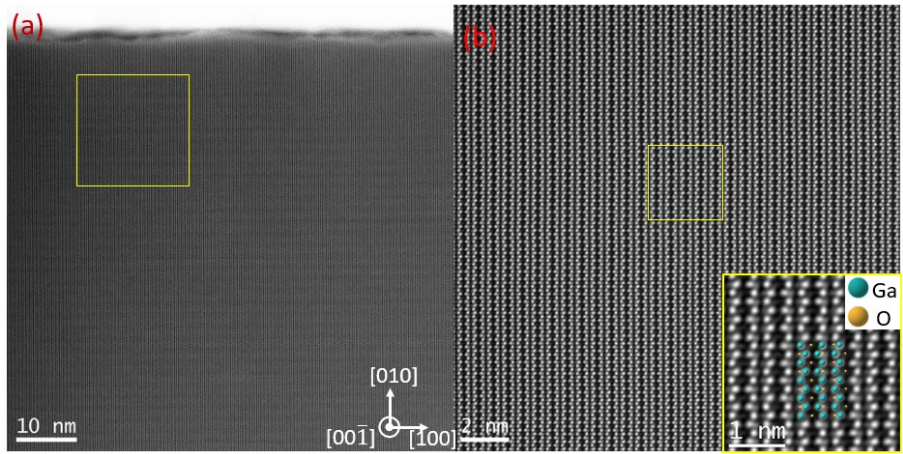

**Fig.4** STEM images of delta-doped β-Ga$_2$O$_3$ films showing high crystalline quality (a) HAADF-STEM image of the sample in the [00$\bar{1}$] projection (b) zoomed in image of the region surrounding the Si delta sheet

STEM (scanning/transmission electron microscopy) imaging is performed on the delta-doped layers to ascertain the effect of growth interruption process (pre-purge and post-purge steps) on the crystalline quality of the as-grown films. High resolution High angle annular dark field-scanning transmission electron microscopy (HAADF-STEM) imaging is carried out using a FEI Titan G2 60-300 transmission electron microscope (TEM) at 300kV. A condenser aperture of 70um was used with a convergence angle of 30 mrads and the annular detector collection angles in the 42-250 mrad range. The probe current is approximately 90 pA. In STEM imaging, the contrast is proportional to the Z number of the atom, i.e. the heavier the atom, the brighter the contrast. Figure 4.(a) is a HAADF-STEM image of the sample in the [00$\bar{1}$] projection. The top of the image corresponds to the Ni contact layer whereas the homogeneous darker contrast corresponds to β-Ga$_2$O$_3$. The yellow square highlights an area that contains the Si δ-Sheet and its corresponding image is shown in Figure 4. (b). The overall contrast of the image is homogeneous with no observable defects that could be potentially caused by the Si δ-Sheet. A zoom-in image superimposed with the β-Ga$_2$O$_3$ model is shown in the insert. Ga and O atoms are represented by teal and yellow spheres respectively. From these observation, it can be seen that the introduction of the Si δ-Sheet into the β-Ga$_2$O$_3$ does not affect its crystal structure. Indeed, the doping level of Si (~1 x 10$^{13}$ cm$^{-2}$) is low enough to not cause any damage in the β-Ga$_2$O$_3$ lattice.

**Table I.** Details of Hall measurements performed on MOVPE-grown β-Ga$_2$O$_3$ films

| Sample | Epitaxial Structure | Sheet charge (cm$^{-2}$) | Equivalent doping density (cm$^{-3}$) | Mobility (cm$^2$/V. s) | Thickness (nm) |
|---|---|---|---|---|---|
| A | 9 coupled delta wells(~1x 10$^{13}$ each well) with 10 nm UID spacer | 9.6 x 10$^{13}$ | 1.1 x 10$^{19}$ | 77 | 85 |
| B | Uniformly doped β-Ga$_2$O$_3$ | 8.5 x 10$^{13}$ | 1 x 10$^{19}$ | 50 | 85 |
| C | Single delta sheet | 6.2 x 10$^{12}$ | - | 83 | - |

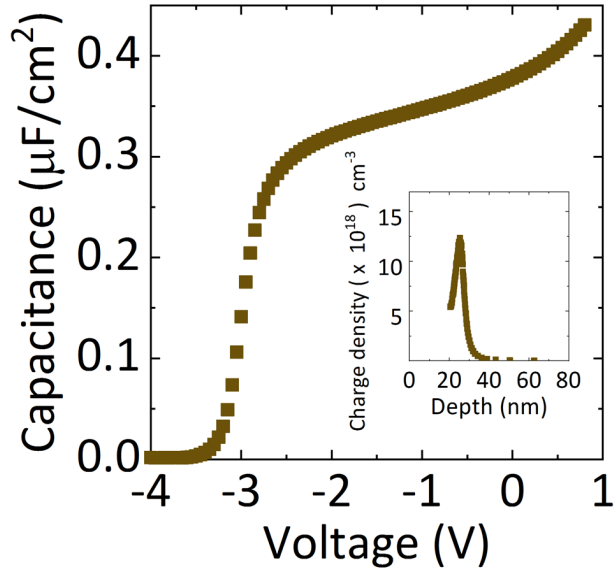

**Fig.5** CV data of the delta sheet (on Fe-doped substrate) showing a sheet charge of 6.1 x 10$^{12}$ cm$^{-2}$ with inset showing the depth profile of the delta sheet

In order to understand the electron transport of the delta doped β-Ga$_2$O$_3$ layers, hall measurements are done on delta-doped films on Fe-doped substrates. Based on theoretical calculations, delta doped films are expected to have reduced scattering compared to a uniformly doped semiconductor[15], due to spread of the electron wavefunction. This can lead to an increase in carrier mobility as compared to a uniformly doped semiconductor. By employing multiple delta doped quantum wells, the mobility of the epitaxial film can be enhanced even further[17]. To understand the effect of dopant profiles on transport properties, hall measurements are performed on three different films – Sample A- 9 delta sheets with a 10 nm UID β-Ga$_2$O$_3$ spacer layer (5 nm cap layer for ohmic contact formation), Sample B – uniformly doped β-Ga$_2$O$_3$, Sample C – single delta sheet β-Ga$_2$O$_3$ with 120 nm buffer layer and

25 nm cap layer All the growths are performed at the similar growth temperature (600 - 650 °C), details of the hall measurements are listed in table 1. The thickness and equivalent charge volume density of samples A and B are kept similar for a fair comparison of carrier mobility. We found that making low resistance direct ohmic contact to a single delta sheet (sample C) is challenging. To circumvent this problem low temperature MOVPE regrowth is performed for direct contact to the electron sheet charge. Regrowth is done by using a $SiO_2$ mask after etching the epitaxial film and regrowing heavily doped (n+) $\beta$-$Ga_2O_3$ contact layers to the delta channel. Further details of the regrowth process will be reported elsewhere. Hall measurements of sample C showed a sheet charge of 6.2 x $10^{12}$ $cm^{-2}$ and a mobility of 83 $cm^2$/V.s. Additionally, we performed CV measurement to characterize the density of the delta sheet (Fig. 5). Equilibrium sheet charge density of 6.1 x $10^{12}$ $cm^{-2}$ is extracted from the CV data. The CV measured charge density agrees well with the hall data, indicating the reliability of the electrical measurements. The measured mobility of the delta-doped films is comparable to MBE-grown $\beta$-$Ga_2O_3$ films reported in literature[16].

Hall measurements of samples A and B showed that the mobility of sample A with coupled delta sheets (77 $cm^2$/V. s) is ~50 % higher compared to uniformly doped $\beta$-$Ga_2O_3$ layers (50 $cm^2$/V.s ). The equivalent doping density of sample A and B is close to 1 x $10^{19}$ $cm^{-3}$. The increase in carrier mobility could be attributed to spread of the electron wavefunction to undoped regions of the film. This lowers the ionized impurity scattering in the films, leading to improvement in overall mobility. However, the enhancement in mobility is lower when compared to similar structures studied in MBE grown GaAs[15]. One potential reason could be relatively large FWHM of Si doped $\beta$-$Ga_2O_3$ compared to Si-doped GaAs reported in literature. By further reducing the FWHM of the Si donors, further mobility improvement could potentially be attained[15].

In summary, we have showed that surface segregation of silicon leads to large FWHM values in MOVPE-grown $\beta$-$Ga_2O_3$ delta doped films. SIMS measurements indicate that the surface segregation coefficient(R) reduced linearly with growth temperature. By reducing the growth temperature to 600 °C, silicon delta sheets with electron concentration FWHM (3.2 nm) close to the theoretical limit are realized. TEM measurement on the delta-doped film did not show any observable degradation in crystalline quality. Hall measurements on coupled delta sheets showed a ~50 % improvement in electron mobility compared to a uniformly-doped film with equivalent 3D charge. Electrical characterization of the single delta sheet showed a sheet charge $n_s$ – 6.2 x $10^{12}$ $cm^{-2}$ and a carrier mobility of 83 $cm^2$/V.s. This work on the demonstration of sharp delta doping in MOVPE-grown $\beta$-$Ga_2O_3$ films show the promise of delta-doped channel layers for high-power and RF applications.

**Acknowledgements:**

This material is based upon work supported by the Air Force Office of Scientific Research under award number FA9550-18-1-0507 and monitored by Dr. Ali Sayir. Any opinions, findings, conclusions or recommendations expressed in this material are those of the authors and do not necessarily reflect the views of the United States Air Force. Praneeth Ranga acknowledges support from University of Utah Graduate Research Fellowship 2020-2021.This work was performed in part at the Utah Nanofab sponsored by the College of Engineering and the Office of the Vice President for Research. The authors thank the Air Force Research Laboratory's Sensors Directorate for their discussions with them. We also thank Prof. Michael Scarpulla at the University of Utah for providing access to equipment used in this work. The electron microscopy work was performed in the Materials Characterization lab (MCL) at the Materials Research Institute (MRI) at the Pennsylvania State University. The work at PSU was supported by the AFOSR program FA9550-18-1-0277 (GAME MURI, Dr. Ali Sayir, Program Manager).

**Data availability statement:**

The data that support the findings of this study are available from the corresponding author upon reasonable request.

References:

[1] M. Higashiwaki, K. Sasaki, A. Kuramata, T. Masui, and S. Yamakoshi, Appl. Phys. Lett. **100**, 013504 (2012).
[2] A. Kuramata, K. Koshi, S. Watanabe, Y. Yamaoka, T. Masui, and S. Yamakoshi, Jpn. J. Appl. Phys. **55**, 1202A2 (2016).
[3] Z. Feng, A.F.M. Anhar Uddin Bhuiyan, M.R. Karim, and H. Zhao, Appl. Phys. Lett. **114**, 250601 (2019).
[4] K. Sasaki, A. Kuramata, T. Masui, E.G. Víllora, K. Shimamura, and S. Yamakoshi, Appl. Phys. Express **5**, 035502 (2012).
[5] M. Higashiwaki, K. Sasaki, H. Murakami, Y. Kumagai, A. Koukitu, A. Kuramata, T. Masui, and S. Yamakoshi, Semicond. Sci. Technol. **31**, 034001 (2016).
[6] S. Rafique, M.R. Karim, J.M. Johnson, J. Hwang, and H. Zhao, Appl. Phys. Lett. **112**, 052104 (2018).
[7] R. Miller, F. Alema, and A. Osinsky, IEEE Trans. Semicond. Manuf. **31**, 467 (2018).
[8] Y. Zhang, F. Alema, A. Mauze, O.S. Koksaldi, R. Miller, A. Osinsky, and J.S. Speck, APL Mater. **7**, 022506 (2019).
[9] K. Goto, K. Konishi, H. Murakami, Y. Kumagai, B. Monemar, M. Higashiwaki, A. Kuramata, and S. Yamakoshi, Thin Solid Films **666**, 182 (2018).
[10] M. Baldini, M. Albrecht, A. Fiedler, K. Irmscher, R. Schewski, and G. Wagner, ECS J. Solid State Sci. Technol. **6**, Q3040 (2017).
[11] S. Bin Anooz, R. Grüneberg, C. Wouters, R. Schewski, M. Albrecht, A. Fiedler, K. Irmscher, Z. Galazka, W. Miller, G. Wagner, J. Schwarzkopf, and A. Popp, Appl. Phys. Lett. **116**, 182106 (2020).
[12] Z. Xia, H. Chandrasekar, W. Moore, C. Wang, A.J. Lee, J. McGlone, N.K. Kalarickal, A. Arehart, S. Ringel, F.

Yang, and S. Rajan, Appl. Phys. Lett. **115**, 252104 (2019).

[13] S. Sharma, K. Zeng, S. Saha, and U. Singisetti, IEEE Electron Device Lett. **41**, 836 (2020).

[14] E.F. Schubert, A. Fischer, and K. Ploog, IEEE Trans. Electron Devices **33**, 625 (1986).

[15] X. Zheng, T.K. Carns, K.L. Wang, and B. Wu, Appl. Phys. Lett. **62**, 504 (1993).

[16] Z. Xia, C. Joishi, S. Krishnamoorthy, S. Bajaj, Y. Zhang, M. Brenner, S. Lodha, and S. Rajan, IEEE Electron Device Lett. **39**, 568 (2018).

[17] S. Krishnamoorthy, Z. Xia, S. Bajaj, M. Brenner, and S. Rajan, Appl. Phys. Express **10**, 051102 (2017).

[18] Z. Xia, H. Xue, C. Joishi, J. Mcglone, N.K. Kalarickal, S.H. Sohel, M. Brenner, A. Arehart, S. Ringel, S. Lodha, W. Lu, and S. Rajan, IEEE Electron Device Lett. **40**, 1052 (2019).

[19] Z. Feng, A.F.M.A.U. Bhuiyan, Z. Xia, W. Moore, Z. Chen, J.F. McGlone, D.R. Daughton, A.R. Arehart, S.A. Ringel, S. Rajan, and H. Zhao, Phys. Status Solidi RRL – Rapid Res. Lett. **14**, 2000145 (2020).

[20] A. Bhattacharyya, P. Ranga, S. Roy, J. Ogle, L. Whittaker-Brooks, and S. Krishnamoorthy, ArXiv Prepr. ArXiv200800303 (2020).

[21] Y. Zhang, A. Neal, Z. Xia, C. Joishi, J.M. Johnson, Y. Zheng, S. Bajaj, M. Brenner, D. Dorsey, K. Chabak, G. Jessen, J. Hwang, S. Mou, J.P. Heremans, and S. Rajan, Appl. Phys. Lett. **112**, 173502 (2018).

[22] A. Kumar and U. Singisetti, ArXiv Prepr. ArXiv200300959 (2020).

[23] A.F.M. Anhar Uddin Bhuiyan, Z. Feng, J.M. Johnson, Z. Chen, H.-L. Huang, J. Hwang, and H. Zhao, Appl. Phys. Lett. **115**, 120602 (2019).

[24] J.B. Varley, A. Perron, V. Lordi, D. Wickramaratne, and J.L. Lyons, Appl. Phys. Lett. **116**, 172104 (2020).

[25] P. Ranga, A. Rishinaramangalam, J. Varley, A. Bhattacharyya, D. Feezell, and S. Krishnamoorthy, Appl. Phys. Express **12**, 111004 (2019).

[26] P. Ranga, A. Bhattacharyya, A. Rishinaramangalam, Y.K. Ooi, M.A. Scarpulla, D. Feezell, and S. Krishnamoorthy, Appl. Phys. Express **13**, 045501 (2020).

[27] G.Y. Zhao, M. Adachi, H. Ishikawa, T. Egawa, M. Umeno, and T. Jimbo, Appl. Phys. Lett. **77**, 2195 (2000).

[28] C. Bayram, J.L. Pau, R. McClintock, and M. Razeghi, J. Appl. Phys. **104**, 083512 (2008).

[29] E.C. Young, N. Grandjean, T.E. Mates, and J.S. Speck, Appl. Phys. Lett. **109**, 212103 (2016).

[30] A.A. Marmalyuk, O.I. Govorkov, A.V. Petrovsky, D.B. Nikitin, A.A. Padalitsa, P.V. Bulaev, I.V. Budkin, and I.D. Zalevsky, J. Cryst. Growth **237**, 264 (2002).

[31] A. Mauze, Y. Zhang, T. Mates, F. Wu, and J.S. Speck, Appl. Phys. Lett. **115**, 052102 (2019).

[32] R. Sharma, M.E. Law, C. Fares, M. Tadjer, F. Ren, A. Kuramata, and S.J. Pearton, AIP Adv. **9**, 085111 (2019).

[33] K. Sasaki, M. Higashiwaki, A. Kuramata, T. Masui, and S. Yamakoshi, Appl. Phys. Express **6**, 086502 (2013).

[34] K. Muraki, S. Fukatsu, Y. Shiraki, and R. Ito, Appl. Phys. Lett. **61**, 557 (1992).

[35] E.F. Schubert, R.F. Kopf, J.M. Kuo, H.S. Luftman, and P.A. Garbinski, Appl. Phys. Lett. **57**, 497 (1990).